\documentclass{article}
\usepackage{spconf,amsmath,graphicx}
\usepackage{multirow}
\usepackage{url}
\usepackage{cite}
\usepackage{subcaption}
\usepackage{cool}
\usepackage{booktabs} 
\usepackage{listings}
\usepackage{url}
\usepackage{listings}
\usepackage{slashbox}
\usepackage{color}
\definecolor{deepblue}{rgb}{0,0,0.5}
\definecolor{deepred}{rgb}{0.6,0,0}
\definecolor{deepgreen}{rgb}{0,0.5,0}

\newcommand\pythonstyle{\lstset{
language=Python,
basicstyle=\ttm,
otherkeywords={self},             
keywordstyle=\ttb\color{deepblue},
emph={MyClass,__init__},          
emphstyle=\ttb\color{deepred},    
stringstyle=\color{deepgreen},
frame=tb,                         
showstringspaces=false            %
}}

\lstnewenvironment{python}[1][]
{
\pythonstyle
\lstset{#1}
}
{}

\newcommand\pythoninline[1]{{\pythonstyle\lstinline!#1!}}

\title{The PyTorch-Kaldi Speech Recognition Toolkit}
\name{Mirco Ravanelli$^1$, Titouan Parcollet$^2$, Yoshua Bengio$^{1*}$}
\address{
  $^1$ Mila, Universit\'e de Montr\'eal , $^*$CIFAR Fellow \\
  $^2$ LIA, Universit\'e d'Avignon}
  
\begin{document}
\ninept
\maketitle
\begin{abstract}
The availability of open-source software is playing a remarkable role in the popularization of speech recognition and deep learning. Kaldi, for instance, is nowadays an established framework used to develop state-of-the-art speech recognizers. PyTorch is used to build neural networks with the Python language and has recently spawn tremendous interest within the machine learning community thanks to its simplicity and flexibility.

The PyTorch-Kaldi project aims to bridge the gap between these popular toolkits, trying to inherit the efficiency of Kaldi and the flexibility of PyTorch. PyTorch-Kaldi is not only a simple interface between these software, but it embeds several useful features for developing modern speech recognizers. For instance, the code is specifically designed to naturally plug-in user-defined acoustic models. As an alternative, users can exploit several pre-implemented neural networks that can be customized using intuitive configuration files. PyTorch-Kaldi  supports multiple feature and label streams as well as combinations of neural networks, enabling the use of complex neural architectures. The toolkit is publicly-released along with a rich documentation and is designed to properly work locally or on HPC clusters.

Experiments, that are conducted on several datasets and tasks, show that PyTorch-Kaldi can effectively be used to develop modern state-of-the-art speech recognizers.


\end{abstract}
\noindent\textbf{Index Terms}: speech recognition, deep learning, Kaldi, PyTorch.

\section{Introduction}
Over the last years, we witnessed a progressive improvement and maturation of Automatic Speech Recognition (ASR) technologies \cite{lideng,ravanelli_thesis}, that have reached unprecedented performance levels and are nowadays used by millions of users worldwide. 

A key role in this technological breakthrough is being played by deep learning \cite{Goodfellow-et-al-2016-Book}, that contributed to overcoming previous speech recognizers based on Gaussian Mixture Models (GMMs).
Beyond deep learning, other factors have played a role in the progress of the field. A number of speech-related projects such as AMI~\cite{ami} and DIRHA~\cite{lrec} and speech recognition challenges such as CHiME~\cite{chime3}, Babel, and Aspire, have remarkably fostered the progress in ASR. The public distribution of large datasets such as Librispeech \cite{librispeech} has also played an important role to establish common evaluation frameworks and tasks. 

Among the others factors, the development of open-source software such as HTK~\cite{htkbook}, Julius \cite{julius}, CMU-Sphinx, RWTH-ASR~\cite{rwth}, LIA-ASR~\cite{lia_asr}  and, more recently, the Kaldi toolkit~\cite{kaldi_short} have further helped popularize ASR, making both research and development of novel ASR applications significantly easier.  

Kaldi currently represents the most popular ASR toolkit. It relies on finite-state transducers (FSTs) \cite{mohri} and provides a set of C++ libraries for efficiently implementing state-of-the-art speech recognition systems. Moreover, the toolkit includes a large set of recipes that cover all the most popular speech corpora. In parallel to the development of this ASR-specific software, several general-purpose deep learning frameworks, such as Theano~\cite{theano}, TensorFlow~\cite{tensorflow}, and CNTK~\cite{CNTK}, have gained popularity in the machine learning community. These toolkits offer a huge flexibility in the neural network design and can be used for a variety of deep learning applications.

PyTorch \cite{PyTorch} is an emerging python package that implements efficient GPU-based tensor computations and facilitates the design of neural architectures, thanks to proper routines for automatic gradient computation. An interesting feature of PyTorch lies in its modern and flexible design, that naturally supports dynamic neural networks. In fact, the computational graph is dynamically constructed on-the-fly at running time rather than being statically compiled.

The PyTorch-Kaldi project aims to bridge the gap between Kaldi and PyTorch\footnote{\url{github.com/mravanelli/pytorch-kaldi/}.}. Our toolkit implements acoustic models in PyTorch, while feature extraction, label/alignment computation, and decoding are performed with Kaldi, making it suitable to develop state-of-the-art DNN-HMM speech recognizers.  PyTorch-Kaldi natively supports several DNNs, CNNs, and RNNs models. 
Combinations between deep learning models, acoustic features, and labels are also supported, enabling the use of complex neural architectures.  For instance, users can employ a cascade between CNNs, LSTMs, and DNNs, or run in parallel several models that share some hidden layers. Users can also explore different acoustic features, context duration, neuron activations (e.g., ReLU, leaky ReLU), normalizations (e.g., batch \cite{batchnorm} and layer normalization \cite{layer_norm}), cost functions, regularization strategies (e.g, L2, dropout \cite{dropout}), optimization algorithms (e.g., Adam \cite{adam}, RMSPROP), and many other hyper-parameters of an ASR system through simple edits of configuration files.

The toolkit is designed to make the integration of user-defined acoustic models as simple as possible. In practice, users can embed their deep learning model and conduct ASR experiments even without being fully familiar with the complex speech recognition pipeline. The toolkit can perform computations on both local machines and HPC cluster, and supports multi-gpu training, recovery strategy, and automatic data chunking.

The experiments, conducted on several datasets and tasks, have shown that PyTorch-Kaldi makes it possible to easily develop competitive state-of-the-art speech recognition systems.


\section{The PyTorch-Kaldi Project}
Some other speech recognition toolkits have been recently developed using the python language. PyKaldi \cite{pykaldi}, for instance, is an easy-to-use Python wrapper for the C++ code of Kaldi and OpenFst libraries. Differently from our toolkit, however, the current version of the PyKaldi does not provide several pre-implemented and ready-to-use neural models.  Another python project is ESPnet \cite{ESPnet}. ESPnet is an end-to-end speech processing toolkit, mainly focuses on end-to-end speech recognition and end-to-end text-to-speech. The main difference with our project is the current version of PyTorch-Kaldi implements hybrid DNN-HMM speech recognizers.

\begin{figure}[t!]
 \centering
   \includegraphics[scale=0.70]{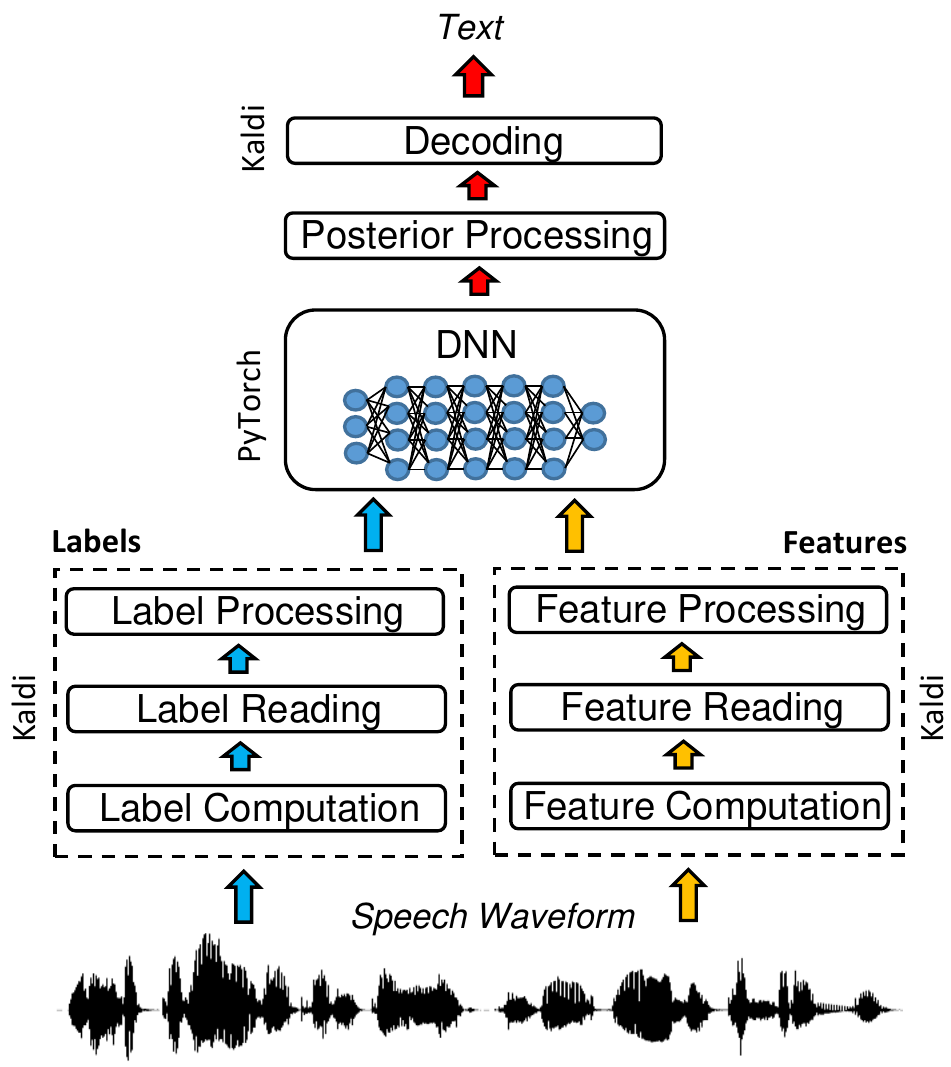}
 \caption{An overview of the PyTorch-Kaldi architecture.}
 \label{fig:arch}
 \end{figure}

An overview of the architecture adopted in PyTorch-Kaldi is reported in Fig. \ref{fig:arch}. 
The main script \textit{$run\_exp.py$} is written in \textit{python} and manages all the phases involved in an ASR system, including feature and label extraction, training, validation, decoding, and scoring.
The toolkit is detailed in the following sub-sections.

\subsection{Configuration file}
The main script takes as input a configuration file in INI format\footnote{The configuration file is fully described in the project documentation.}, that is composed of several sections.
The section $[Exp]$ specifies some high-level information such as the folder used for the experiment, the number of training epochs, the random seed. It also allows users to specify whether the experiments have to be conducted on a CPU, GPU, or on multiple GPUs.
The configuration file continues with the $[dataset*]$ sections, that specify information on features and labels, including the paths where they are stored, the characteristics of the context window \cite{acw_sc}, and the number of chunks in which the speech dataset must be split. 
The neural models are described in the $[architecture*]$ sections, while the $[model]$ section defines how these neural networks are combined. 
The latter section exploits a simple meta-language that is automatically interpreted by the $run\_exp.py$ script.
Finally, the configuration file defines the decoding parameters in the $[decoding]$ section.

\subsection{Features}
The feature extraction is performed with Kaldi, that natively provides c++ libraries (e.g., \textit{compute-mfcc-feats}, \textit{compute-fbank-feats}, \textit{compute-plp-feats}) to efficiently extract the most popular speech recognition features.  
The computed coefficients are stored in binary archives (with extension \textit{.ark}) and are later imported into the python environment using the \textit{kaldi-io} utilities inherited from the kaldi-io-for-python project\footnote{\url{github.com/vesis84/kaldi-io-for-python}}. 
The features are then processed by the function \textit{load-chunk},  that performs context window composition, shuffling, as well as mean and variance normalization. As outlined before,  PyTorch-Kaldi can manage multiple feature streams. For instance, users can define models that exploit combinations of MFCCs, FBANKs, PLP, and fMLLR \cite{mllr} coefficients.

\subsection{Labels}
The main labels used for training the acoustic model derive from a forced alignment procedure between the speech features and the sequence of context-dependent phone states computed by Kaldi with a phonetic decision tree. To enable multi-task learning, PyTorch-Kaldi supports multiple labels. For instance, it is possible to jointly load both context-dependent and context-independent targets and use the latter ones to perform monophone regularization \cite{ravanelli15,mono_reg}. It is also possible to employ models based on an ecosystem of neural networks performing different tasks, as done in the context of joint training between speech enhancement and speech recognition \cite{ravanelli_SLT,joint2} or in the context of the recently-proposed cooperative networks of deep neural networks \cite{ravanelli_icassp}.

\subsection{Chunk and Mini-batch Composition}
PyTorch-Kaldi automatically splits the full dataset into a number of chunks, which are composed of labels and features randomly sampled from the full corpus. Each chunk is then stored into the GPU or CPU memory and processed by the neural training algorithm $run\_nn.py$. The toolkit dynamically composes different chunks at each epoch.
A set of mini-batches are then derived from them. Mini-batches are composed of few  training examples that are used for gradient computation and parameter optimization. 

The way mini-batches are gathered strongly depends on the typology of the neural network. 
For feed-forward models, the mini-batches are composed of randomly shuffled features and labels sampled from the chunk. For recurrent networks, the minibatches must be composed of full sentences. Different sentences, however, are likely to have different duration, making zero-padding necessary to form mini-batches of the same size. PyTorch-Kaldi sorts the speech sequences in ascending order according to their lengths (i.e., short sentences are processed first). This approach minimizes the need of zero-paddings and turned out to be helpful to avoid possible biases on batch normalization statistics. Moreover, it has been shown useful to slightly boost the performance and to improve the numerical stability of gradients. 

\subsection{DNN acoustic modeling}
Each minibatch is processed by a neural network implemented with PyTorch, that takes as input the features and as outputs a set of posterior probabilities over the context-dependent phone states. The code is designed to easily plug-in customized models. As reported in the pseudo-code reported in Fig. \ref{fig:code}, the new model can be simply defined by adding a new class into the \textit{neural\_nets.py}. The class must be composed of an initialization method, that specifies the parameters with their initialization, and a forward method that defines the computations to perform. 

\begin{figure}[t!]
\caption{Adding a user model into PyTorch-Kaldi.}
\lstset{language=Python}
\lstset{frame=lines}
\lstset{label={lst:code_direct}}
\lstset{basicstyle=\footnotesize}
\begin{lstlisting}
class my_NN(nn.Module):
    def __init__(self, options):
        super(my_NN, self).__init__()
        # Definition of Model Parameters
        # Parameter Initialization
        
    def forward(self, minibatch):
        # Definition of Model Computations
        return [output_prob]
\end{lstlisting}
\label{fig:code}
\end{figure}

As an alternative, a number of pre-defined state-of-the-art neural models are natively implemented within the toolkit. The current version supports standard MLPs, CNNs, RNNs, LSTM, and GRU models. Moreover, it supports some advanced recurrent architectures, such as the recently-proposed Light GRU \cite{li_gru} and twin-regularized RNNs \cite{twin_reg}. The SincNet model \cite{SincNet,sincnet_irasl} is also implemented to perform speech recognition from raw waveform directly. The hyperparameters of the model (such as learning rate, number of neurons, number of layers, dropout factor, etc.) can be tuned using a utility that implements the random search algorithm \cite{random_search}.

\subsection{Decoding and Scoring}
The acoustic posterior probabilities generated by the neural network are normalized by their prior before feeding the HMM-based decoder of Kaldi. The decoder merges the acoustic scores with the language probabilities derived by an n-gram language model and tries to retrieve the sequence of words uttered in the speech signal using a beam-search algorithm. The final Word-Error-Rate (WER) score is computed with the NIST SCTK scoring toolkit.

\section{Experimental Setup} \label{sec:setup}
In the following sub-sections, the corpora, and the DNN setting adopted for the experimental activity are described.

\subsection{Corpora and Tasks}
The first set of experiments was performed with the TIMIT corpus, considering the standard phoneme recognition task (aligned with the Kaldi s5 recipe \cite{kaldi_short}). 

To validate our model in a more challenging scenario, experiments were also conducted in distant-talking conditions with the DIRHA-English dataset\footnote{This dataset is distributed by the Linguistic Data Consortium (LDC).} \cite{dirha_asru,rav_is16}. 
Training was based on the original WSJ-5k corpus (consisting of $7,138$ sentences uttered by $83$ speakers) that was contaminated with a set of impulse responses measured in a domestic environment \cite{rav_is16}.
The test phase was carried out with the real part of the dataset, consisting of $409$ WSJ sentences uttered in the aforementioned environment by six native American speakers. 

Additional experiments were conducted with the CHiME~4 dataset \cite{chime3}, that is based on speech data recorded in four noisy environments (on a bus, cafe, pedestrian area, and street junction). The training set is composed of $43,690$ noisy WSJ sentences recorded by five microphones (arranged on a tablet) and uttered by a total of $87$ speakers. 
The test set \textit{ET-real} considered in this work is based on $1,320$ real sentences uttered by four speakers, while the subset \textit{DT-real} has been used for hyperparameter tuning. The CHiME experiments were based on the single channel setting \cite{chime3}. 

Finally, experiments were performed with the LibriSpeech~\cite{librispeech} dataset. 
We used the training subset composed of $100$ hours and the \textit{dev-clean} set for the hyperparameter search. Test results are reported on the \textit{test-clean} part using the \textit{fglarge} decoding graph inherited from the Kaldi s5 recipe. 

\subsection{DNN setting}
The experiments consider different acoustic features, i.e., $39$ MFCCs ($13$ static+$\Delta$+$\Delta\Delta$), $40$ log-mel filter-bank features (FBANKS), as well as $40$ fMLLR features \cite{mllr} (extracted as reported in the s5 recipe of Kaldi), that were computed using windows of $25$ ms with an overlap of $10$ ms.
The feed-forward models were initialized according to the \textit{Glorot}'s scheme \cite{xavier}, while recurrent weights were initialized with orthogonal matrices \cite{orth_init}. 
Recurrent dropout was used as a regularization technique \cite{drop_asru}. 
Batch normalization was adopted for feed-forward connections only, as proposed in \cite{ravanelli_is17}.
The optimization was done using the RMSprop algorithm running for $24$ epochs. The performance on the development set was monitored after each epoch and the learning rate was halved when the relative performance improvement went below~$0.1\%$.  
The main hyperparameters of the model (i.e., learning rate, number of hidden layers, hidden neurons per layer, dropout factor, as well as the twin regularization term $\lambda$) were tuned on the development datasets. 

\section{Baselines}
In this section, we discuss the baselines obtained with TIMIT, DIRHA, CHiME, and LibriSpeech datasets. As a showcase to illustrate the main functionalities of the PyTorch-Kaldi toolkit, we first report the experimental validation conducted on TIMIT.

Table \ref{tab:res1} shows the performance obtained with several feed-forward and recurrent models using different features. To ensure a more accurate comparison between the architectures, five experiments varying the initialization seeds were conducted for each 
model and feature. The table thus reports the average \textit{phone error rates} (PER)\footnote{Standard deviations range between $0.15$ and $0.2$ for all the experiments.}.
\begin{table}[t!]
\caption{PER(\%) obtained for the test set of TIMIT with various neural architectures.}
\centering

\begin{tabular}{l|cccc}  
    & MFCC &  FBANK & fMLLR \\ \hline
MLP & 18.2  & 18.7 & 16.7   \\ 
RNN & 17.7  & 17.2  & 15.9   \\ 
LSTM & 15.1  & 14.3  & 14.5   \\ 
GRU & 16.0  & 15.2  & 14.9   \\ 
Li-GRU & 15.3 & 14.9 & \textbf{14.2} \\
\bottomrule
\end{tabular}
\label{tab:res1}
\end{table}
Results show that, as expected, fMLLR features outperform MFCCs and FBANKs coefficients, thanks to the speaker adaptation process. Recurrent models significantly outperform the standard MLP one, especially when using LSTM, GRU, and Li-GRU architecture, that effectively address gradient vanishing through multiplicative gates. The best result (PER=$14.2$\%) is obtained with the Li-GRU model \cite{li_gru}, that is based on a single gate and thus saves $33$\% of the computations over a standard GRU.

Table \ref{tab:res2} details the impact of some popular techniques implemented in PyTorch-Kaldi for improving the ASR performance.
\begin{table}[t!]
\caption{PER(\%) obtained on TIMIT when progressively applying some techniques implemented within PyTorch-Kaldi.}
\centering

\begin{tabular}{l|ccccc}  
    & RNN & LSTM & GRU & Li-GRU  \\ \hline
Baseline & 16.5  &  16.0 & 16.6 & 16.3 \\ 
+ Incr. Seq. length & 16.6  &  15.3 & 16.1 & 15.4 \\ 
+ Recurrent Dropout & 16.4  &  	15.1 & 15.4 & 14.5 \\ 
+ Batch Normalization & 16.0  &  14.8 & 15.3 & 14.4 \\ 
+ Monophone Reg. & 15.9  &  14.5 & 14.9 & \textbf{14.2 }\\ 
\bottomrule
\end{tabular}
\label{tab:res2}
\end{table}
The first row (Baseline) reports the performance achieved with a basic recurrent model, where powerful techniques such as dropout and batch normalization are not adopted. The second row highlights the performance gain that is achieved when progressively increasing the sequence length during training. In this case, we started the training by truncating the speech sentence at $100$ steps (i.e, approximately 1 second of speech) and we progressively double the maximum sequence duration at every epoch. This simple strategy generally improves the system performance since it encourages the model to first focus on short-term dependencies and learn longer-term ones only at a later stage. The third row shows the improvement achieved when adding recurrent dropout. Similarly to \cite{drop_asru,ravanelli_is17}, we applied the same dropout mask for all the time steps to avoid gradient vanishing problems. The fourth line, instead, shows the benefits derived from batch normalization \cite{batchnorm}. 
Finally, the last line shows the performance achieved when also applying monophone regularization \cite{mono_reg}. In this case, we employ a multi-task learning strategy by means of two softmax classifiers: the first one estimates context-dependent states, while the second one predicts monophone targets. As observed in \cite{mono_reg}, our results confirm that this technique can successfully be used as an effective regularizer.

The experiments discussed so far are based on single neural models. In Table \ref{tab:res3} we compare our best Li-GRU system with a more complex architecture based on a combination of feed-forward and recurrent models fed by a concatenation of features. To the best of our knowledge, the PER=$13.8$\% achieved by the latter system yields the best-published performance on the TIMIT test-set. 
\begin{table}[t!]
\caption{PER(\%) obtained by combining multiple neural networks and acoustic features.}
\centering

\begin{tabular}{l|lc}  
Architecture         & Features & PER (\%) \\ \hline
Li-GRU       &    fMLLR     &    14.2  \\ 
MLP+Li-GRU+MLP &    MFCC+FBANK+fMLLR    &   \textbf{13.8}     \\ 
\bottomrule
\end{tabular}
\label{tab:res3}
\end{table}

Previous achievements were based on features computed with Kaldi. However, within PyTorch-Kaldi users can employ their own features. Table \ref{tab:res4} shows the results achieved with convolutional models fed by standard FBANKs coefficients or by the raw waveform directly.
\begin{table}[t!]
\caption{PER(\%) obtained with standard convolutional and with the SincNet architectures.}
\centering

\begin{tabular}{l|lc}  
Model         & Features & PER (\%) \\ \hline
CNN       &    FBANK     &    18.3  \\ 
CNN &    Raw waveform   &   18.1     \\ 
SincNet &    Raw waveform    &   \textbf{17.2 }     \\
\bottomrule
\end{tabular}
\label{tab:res4}
\end{table}
The standard CNN based on raw samples performs similarly to the one fed by FBANK features. A performance improvement is observed with SincNet \cite{SincNet}, whose effectiveness in speech recognition is here highlighted for the first time.


We now extend our experimental validation to other datasets. 
With this regard, Table \ref{tab:res5} shows the performance achieved on DIRHA, CHiME, and Librispeech ($100$h) datasets.
\begin{table}[t!]
\caption{WER(\%) obtained for the DIRHA, CHiME, and LibriSpeech ($100$h) datasets with various neural architectures.}
\centering

\begin{tabular}{l|cccc}  
    & DIRHA &  CHiME & LibriSpeech \\ \hline
MLP & 26.1  & 18.7  & 6.5  \\ 
LSTM & 24.8  & 15.5 & 6.4    \\ 
GRU & 24.8  & 15.2 & 6.3   \\ 
Li-GRU & \textbf{23.9} & \textbf{14.6} & \textbf{6.2} \\
\bottomrule
\end{tabular}
\label{tab:res5}
\end{table}
The Table consistently shows better performance with the Li-GRU model, confirming our previous achievements on TIMIT. The results on DIRHA and CHiME show the effectiveness of the proposed toolkit also in noisy condition. To give a comparison, the best Kaldi baseline proposed in $egs/chime4/s5\_1ch$ has a WER(\%)=18.1\%. An end-to-end system trained with ESPnet reaches a WER(\%)=44.99\%, confirming how critical is end-to-end speech recognition is challenging acoustic conditions. 
DIRHA represents another very challenging task, that is characterized by the presence of considerable levels of noise and reverberation.  The WER=$23.9$\% obtained on this dataset represents the best performance published so-far on the single-microphone task. Finally, the performance obtained with Librispeech outperforms the corresponding \textit{p-norm} Kaldi baseline ($WER=6.5\%$) on the considered $100$ hours subset.

\section{Conclusions}
This paper described the PyTorch-Kaldi project, a new initiative that aims to bridge the gap between Kaldi and PyTorch. The toolkit is designed to make the development of an ASR system simpler and more flexible, allowing users to easily plug-in their customized acoustic models. PyTorch-Kaldi also supports combinations of neural architectures, features, and labels, allowing users to possibly employ complex ASR pipelines. The experiments have confirmed that PyTorch-Kaldi can achieve state-of-the-art results in some popular speech recognition tasks and datasets.

The current version of the PyTorch-Kaldi is already publicly-available along with a detailed documentation. The project is still in its initial phase and we invite all potential contributors to participate in it. We hope to build a community of developers larger enough to progressively maintain, improve, and expand the functionalities of our current toolkit. In the future, we plan to increase the number of pre-implemented models, support neural language model training/rescoring, sequence discriminative training, online speech recognition, as well end-to-end training.

\section{Acknowledgment}
We would like to thank Maurizio Omologo, Enzo Telk, and Antonio Mazzaldi for their helpful comments.
This research was enabled in part by support provided by Calcul Qu\'ebec and Compute Canada.

\bibliographystyle{IEEEbib}
\bibliography{mybib}

\begin{thebibliography}{10}

\bibitem{lideng}
D.~Yu and L.~Deng,
\newblock {\em Automatic Speech Recognition -- A Deep Learning Approach},
\newblock Springer, 2015.

\bibitem{ravanelli_thesis}
M.~Ravanelli,
\newblock {\em Deep learning for Distant Speech Recognition},
\newblock PhD Thesis, Unitn, 2017.

\bibitem{Goodfellow-et-al-2016-Book}
I.~Goodfellow, Y.~Bengio, and A.~Courville,
\newblock {\em Deep Learning},
\newblock MIT Press, 2016.

\bibitem{ami}
S.~Renals, T.~Hain, and H.~Bourlard,
\newblock ``{Interpretation of Multiparty Meetings the AMI and Amida
  Projects},''
\newblock in {\em Proc. of HSCMA}, 2008, pp. 115--118.

\bibitem{lrec}
L.~Cristoforetti, M.~Ravanelli, M.~Omologo, A.~Sosi, A.~Abad, M.~Hagmueller,
  and P.~Maragos,
\newblock ``The {DIRHA} simulated corpus,''
\newblock in {\em Proc. of LREC}, 2014, pp. 2629--2634.

\bibitem{chime3}
J.~Barker, R.~Marxer, E.~Vincent, and S.~Watanabe,
\newblock ``{The third CHiME Speech Separation and Recognition Challenge:
  Dataset, task and baselines},''
\newblock in {\em Proc. of ASRU}, 2015, pp. 504--511.

\bibitem{librispeech}
V.~Panayotov, G.~Chen, D.~Povey, and S.~Khudanpur,
\newblock ``{Librispeech: An ASR corpus based on public domain audio books},''
\newblock in {\em Proc. of ICASSP}, 2015, pp. 5206--5210.

\bibitem{htkbook}
S.~Young et~al.,
\newblock {\em {HTK} -- Hidden {M}arkov Model Toolkit}, 2006.

\bibitem{julius}
A.~Lee and T.~Kawahara.,
\newblock ``Recent development of open-source speech recognition engine
  julius,''
\newblock in {\em Proc. of APSIPA-ASC}, 2008.

\bibitem{rwth}
D.~Rybach, S.~Hahn, P.~Lehnen, D.~Nolden, M.~Sundermeyer, Z.~T{\"u}ske,
  S.~Wiesler, R.~Schl{\"u}ter, and H.~Ney,
\newblock ``{RASR - The RWTH Aachen University Open Source Speech Recognition
  Toolkit},''
\newblock in {\em Proc. of ASRU}, 2011.

\bibitem{lia_asr}
G.~Linar{\`e}s, P.~Nocera, D.~Massoni{\'e}, and D.~Matrouf,
\newblock ``The lia speech recognition system: From 10xrt to 1xrt,''
\newblock in {\em Text, Speech and Dialogue}, V{\'a}clav Matou{\v{s}}ek and
  Pavel Mautner, Eds. 2007, pp. 302--308, Springer Berlin Heidelberg.

\bibitem{kaldi_short}
D.~Povey et~al.,
\newblock ``{The Kaldi Speech Recognition Toolkit},''
\newblock in {\em Proc. of ASRU}, 2011.

\bibitem{mohri}
M.~Mohri,
\newblock ``Finite-state transducers in language and speech processing,''
\newblock {\em Computational Linguistics}, vol. 23, no. 2, pp. 269--311, 1997.

\bibitem{theano}
{Theano Development Team},
\newblock ``{Theano: A {Python} framework for fast computation of mathematical
  expressions},''
\newblock {\em arXiv e-prints}, vol. abs/1605.02688, May 2016.

\bibitem{tensorflow}
M.~Abadi et~al.,
\newblock ``Tensorflow: A system for large-scale machine learning,''
\newblock in {\em Proc. of USENIX-OSDI Symposium}, 2016, pp. 265--283.

\bibitem{CNTK}
F.~Seide and A.~Agarwal,
\newblock ``{CNTK: Microsoft's Open-Source Deep-Learning Toolkit},''
\newblock in {\em Proceedings of ACM SIGKDD}, 2016, pp. 2135--2135.

\bibitem{PyTorch}
A.~Paszke et~al.,
\newblock ``Automatic differentiation in pytorch,''
\newblock 2017.

\bibitem{batchnorm}
S.~Ioffe and C.~Szegedy,
\newblock ``Batch normalization: Accelerating deep network training by reducing
  internal covariate shift,''
\newblock in {\em Proc. of ICML}, 2015, pp. 448--456.

\bibitem{layer_norm}
L.~J. Ba, R.~K., and G.~E. Hinton,
\newblock ``Layer normalization,''
\newblock {\em CoRR}, vol. abs/1607.06450, 2016.

\bibitem{dropout}
N.~Srivastava, G.~Hinton, A.~Krizhevsky, I.~Sutskever, and R.~Salakhutdinov,
\newblock ``Dropout: A simple way to prevent neural networks from
  overfitting,''
\newblock {\em Journal of Machine Learning Research}, vol. 15, pp. 1929--1958,
  2014.

\bibitem{adam}
D.P. Kingma and J.~Ba,
\newblock ``Adam: A method for stochastic optimization,''
\newblock in {\em Proc. of ICLR}, 2015.

\bibitem{pykaldi}
D.~Can, V.~R. Martinez, P.~Papadopoulos, and S.~S. Narayanan,
\newblock ``Pykaldi: A python wrapper for kaldi,''
\newblock in {\em Proc. of ICASSP}, 2018.

\bibitem{ESPnet}
S.~{Watanabe}, T.~{Hori}, S.~{Kim}, J.~R. {Hershey}, and T.~{Hayashi},
\newblock ``Hybrid ctc/attention architecture for end-to-end speech
  recognition,''
\newblock {\em IEEE Journal of Selected Topics in Signal Processing}, vol. 11,
  no. 8, pp. 1240--1253, 2017.

\bibitem{acw_sc}
M.~Ravanelli and M.~Omologo,
\newblock ``Automatic context window composition for distant speech
  recognition,''
\newblock {\em Speech Communication}, vol. 101, pp. 34 -- 44, 2018.

\bibitem{mllr}
M.J.F. Gales,
\newblock ``{Maximum Likelihood Linear Transformations for HMM-Based Speech
  Recognition},''
\newblock {\em Computer Speech and Language}, vol. 12, no. 4, pp. 75--98, 1998.

\bibitem{ravanelli15}
M.~Ravanelli and M.~Omologo,
\newblock ``{Contaminated speech training methods for robust DNN-HMM distant
  speech recognition},''
\newblock in {\em Proc. of Interspeech}, 2015, pp. 756--760.

\bibitem{mono_reg}
P.~Bell, P.~Swietojanski, and S.~Renals,
\newblock ``Multitask learning of context-dependent targets in deep neural
  network acoustic models,''
\newblock {\em {IEEE/ACM} Trans. Audio, Speech {\&} Language Processing}, vol.
  25, no. 2, pp. 238--247, 2017.

\bibitem{ravanelli_SLT}
M.~Ravanelli, P.~Brakel, M.~Omologo, and Y.~Bengio,
\newblock ``Batch-normalized joint training for dnn-based distant speech
  recognition,''
\newblock in {\em Proc. of SLT}, 2016, pp. 28--34.

\bibitem{joint2}
A.~Narayanan and D.~Wang,
\newblock ``Joint noise adaptive training for robust automatic speech
  recognition,''
\newblock in {\em Proc. of ICASSP}, 2014, pp. 4380--4384.

\bibitem{ravanelli_icassp}
M.~Ravanelli, P.~Brakel, M.~Omologo, and Y.~Bengio,
\newblock ``A network of deep neural networks for distant speech recognition,''
\newblock in {\em Proc. of ICASSP}, 2017, pp. 4880--4884.

\bibitem{li_gru}
M.~Ravanelli, P.~Brakel, M.~Omologo, and Y.~Bengio,
\newblock ``Light gated recurrent units for speech recognition,''
\newblock {\em IEEE Transactions on Emerging Topics in Computational
  Intelligence}, vol. 2, no. 2, pp. 92--102, April 2018.

\bibitem{twin_reg}
M.~Ravanelli, D.~Serdyuk, and Y.~Bengio,
\newblock ``Twin regularization for online speech recognition,''
\newblock in {\em Proc. of Interspeech}, 2018.

\bibitem{SincNet}
M.~Ravanelli and Y.Bengio,
\newblock ``{Speaker Recognition from raw waveform with SincNet},''
\newblock in {\em Proc. of SLT}, 2018.

\bibitem{sincnet_irasl}
M.~Ravanelli and Y.Bengio,
\newblock ``{Interpretable Convolutional Filters with SincNet},''
\newblock in {\em Proc. of NIPS@IRASL}, 2018.

\bibitem{random_search}
J.~Bergstra and Y.~Bengio,
\newblock ``Random search for hyper-parameter optimization,''
\newblock {\em Journal of Machine Learning Research}, vol. 13, pp. 281--305,
  2012.

\bibitem{dirha_asru}
M.~Ravanelli, L.~Cristoforetti, R.~Gretter, M.~Pellin, A.~Sosi, and M.~Omologo,
\newblock ``{The DIRHA-English corpus and related tasks for distant-speech
  recognition in domestic environments},''
\newblock in {\em Proc. of ASRU}, 2015, pp. 275--282.

\bibitem{rav_is16}
M.~Ravanelli, P.~Svaizer, and M.~Omologo,
\newblock ``Realistic multi-microphone data simulation for distant speech
  recognition,''
\newblock in {\em Proc. of Interspeech}, 2016, pp. 2786--2790.

\bibitem{xavier}
X.~Glorot and Y.~Bengio,
\newblock ``Understanding the difficulty of training deep feedforward neural
  networks,''
\newblock in {\em Proc. of AISTATS}, 2010, pp. 249--256.

\bibitem{orth_init}
Q.V. Le, N.~Jaitly, and G.E. Hinton,
\newblock ``A simple way to initialize recurrent networks of rectified linear
  units,''
\newblock {\em arXiv:1504.00941}, 2015.

\bibitem{drop_asru}
T.~Moon, H.~Choi, H.~Lee, and I.~Song,
\newblock ``{RNNDROP: A novel dropout for RNNS in ASR},''
\newblock in {\em Proc. of ASRU}, 2015.

\bibitem{ravanelli_is17}
M.~Ravanelli, P.~Brakel, M.~Omologo, and Y.~Bengio,
\newblock ``Improving speech recognition by revising gated recurrent units,''
\newblock in {\em Proc. of Interspeech}, 2017.

\end{thebibliography}


\end{document}